\documentclass[prd,twocolumn,
notitlepage,
nofootinbib,aps,tightenlines,
preprintnumbers,amsmath,amssymb,amsfonts,showpacs,superscriptaddress]{revtex4-2}

\usepackage[hyperindex,breaklinks]{hyperref}
\usepackage{graphicx,epstopdf}
\usepackage[utf8]{inputenc}
\usepackage{epsf, array, color}
\usepackage{graphicx}
\usepackage{subfigure}
\usepackage{bbold}
\usepackage{slashed}
\usepackage{amsthm}
\usepackage{amsmath}
\usepackage{comment}
\usepackage[normalem]{ulem}
\usepackage{orcidlink}

\def\eqref#1{Eq.~(\ref{#1})}
\usepackage{natbib,bm} 
\usepackage{breakurl}
\usepackage{multirow}
\usepackage{bbold}
\usepackage{listings}

\DeclareMathAlphabet{\mathcalligra}{T1}{calligra}{m}{n}

\DeclareMathAlphabet{\mathcalligra}{T1}{calligra}{m}{n}

\newcommand{\bea}{\begin{eqnarray}}
\newcommand{\eea}{\end{eqnarray}}

\def\bal#1\eal{\begin{align}#1\end{align}}

\def\Rc{R_{\rm c}}
\def\Mc{M_{\rm c}}

\begin{document}
\title{Probing Heavy Dark Matter in Red Giants}

\author{Sougata Ganguly\orcidlink{0000-0002-8742-0870}}
\email{sganguly0205@ibs.re.kr}
\affiliation{Particle Theory and Cosmology Group, Center for Theoretical Physics of the Universe,
Institute for Basic Science (IBS), Daejeon, 34126, Korea}

\author{Minxi He\orcidlink{0000-0002-2480-6848}}
\email{heminxi@ibs.re.kr}
\affiliation{Particle Theory and Cosmology Group, Center for Theoretical Physics of the Universe,
Institute for Basic Science (IBS), Daejeon, 34126, Korea}

\author{Chang Sub Shin\orcidlink{0000-0002-4211-5653}}
\email{csshin@cnu.ac.kr}
\affiliation{Department of Physics and Institute for Sciences of the Universe,
Chungnam National University, Daejeon 34134, Korea}
\affiliation{Particle Theory and Cosmology Group, Center for Theoretical Physics of the Universe,
Institute for Basic Science (IBS), Daejeon, 34126, Korea}
\affiliation{School of Physics, Korea Institute for Advanced Study, Seoul, 02455, Republic of Korea}

\author{Oscar Straniero\orcidlink{0000-0002-5514-6125}}
\email{oscar.straniero@inaf.it}
\affiliation{INAF, Osservatorio Astronomico d’Abruzzo, 64100 Teramo, Italy}
\affiliation{Istituto Nazionale di Fisica Nucleare - Sezione di Roma, Piazzale Aldo Moro 2, 00185 Roma, Italy}

\author{Seokhoon Yun\orcidlink{0000-0002-7960-3933}}
\email{seokhoon.yun@knu.ac.kr}
\affiliation{Department of Physics, Kyungpook National University, Daegu 41566, Korea}
\affiliation{Particle Theory and Cosmology Group, Center for Theoretical Physics of the Universe,
Institute for Basic Science (IBS), Daejeon, 34126, Korea}

\preprint{CTPU-PTC-25-36}

\begin{abstract}
Red giants (RGs) provide a promising astrophysical environment for capturing dark matter (DM) via elastic scattering with stellar nuclei.
Captured DM particles migrate toward the helium-rich core and accumulate into a compact configuration.
As the DM population grows, it can become self-gravitating and undergo gravitational collapse, leading to adiabatic contraction through interactions with the ambient medium.
The resulting energy release, through elastic scattering and, where relevant, DM annihilation during collapse, locally heats the stellar core and can trigger helium ignition earlier than that predicted by standard stellar evolution.
We analyze the conditions under which DM-induced heating leads to runaway helium burning and identify the critical DM mass required for ignition.
Imposing the observational constraint that helium ignition must not occur before the observed luminosity at the tip of the RG branch, we translate these conditions into bounds on DM properties.
Remarkably, we find that RGs are sensitive to DM, particularly with masses around $10^{11} \,{\rm GeV}$ and spin-independent scattering cross sections near $10^{-37}\,{\rm cm}^2$, which is comparable to the reach of current terrestrial direct detection experiments. Noteworthy, observations of RG stars provide a unique probe for high-mass and large-cross-section DM, a regime that remains currently inaccessible to direct detection experiments.
\end{abstract}

\maketitle

\section{Introduction}
\label{sec:Intro}

Dark matter (DM) constitutes approximately 27\% of the present-day mass–energy content of the Universe, as supported by a wide range of cosmological and astrophysical observations, including the Cosmic Microwave Background (CMB)~\cite{Planck:2018vyg}, galactic rotation curves~\cite{Sofue:2000jx}, and the Bullet Cluster~\cite{Harvey:2015hha}.
Despite its gravitational influence being well established, the fundamental nature of DM remains elusive, largely due to its lack of significant interactions with ordinary matter beyond gravity~\cite{Schumann:2019eaa}.

In the standard $\Lambda$ cold dark matter ($\Lambda$CDM) paradigm, DM is assumed to be cold and collisionless, successfully explaining large-scale structure formation and other gravitational phenomena.
A well-studied candidate in this framework is the weakly interacting massive particle (WIMP), which is thermally produced in the early Universe with a weak scale mass and a weak interaction strength~\cite{Lee:1977ua}.
This scenario naturally accounts for the observed DM abundance, which makes it highly appealing from a theoretical perspective~\cite{Bertone:2016nfn,Billard:2021uyg,Akerib:2022ort}.
However, null results from direct detection experiments such as XENON1T~\cite{XENON:2017vdw} and LUX~\cite{LZ:2022lsv} have excluded much of the WIMP parameter space.
For instance, spin-independent cross sections larger than $\sim 10^{-46}\,\mathrm{cm}^2$ are ruled out for WIMP masses between $5\,{\rm GeV}$ and $100\,\mathrm{GeV}$, well beyond what is required by relic abundance considerations.

These constraints have renewed interest in heavy dark matter (HDM) scenarios, where the DM mass lies well above the weak scale.
The number of recoil events in direct detection experiments decreases with increasing DM mass, which allows HDM to evade current bounds.
However, HDM is typically incompatible with thermal production, as unitarity bounds on annihilation cross sections limit the mass of thermally produced DM to below $\sim 100\,\mathrm{TeV}$~\cite{Griest:1989wd}.
Instead, the correct HDM relic abundance can be achieved through various mechanisms involving sizable interactions with the Standard Model (SM) plasma~\cite{Berlin:2016gtr, Kim:2019udq, Berlin:2017ife, Kramer:2020sbb,Frumkin:2022ror, Chway:2019kft, Baker:2019ndr, Asadi:2021yml, Lennon:2017tqq, Morrison:2018xla, Hooper:2019gtx, Gehrman:2023qjn}. 
These interactions may also lead to observable consequences in current or future experiments.

Beyond laboratory searches, stellar environments offer a promising avenue to probe DM interactions with the visible sector governed by the SM~\cite{Raffelt:1996wa}.
Many studies have demonstrated that stars can constrain DM models through capture~\cite{Press:1985ug,Gould:1987ir} and then energy transport~\cite{1986ApJ...306..703G,Scott:2008ns,Casanellas:2015uga,Lopes:2019jca,Raen:2020qvn} or heating effects, with the sensitivity depending on stellar structure and evolutionary stage. 
Also, DM accumulation can affect the inner structures of main sequence stars, which can constrain asymmetric DM~\cite{Casanellas:2010he,Casanellas:2012jp,Casanellas:2009dp,Rato:2021tfc}. 
Among these, red giants (RGs) are particularly interesting due to the high densities and temperatures in their helium-rich cores.

During the main sequence, hydrogen fusion in the stellar core gradually builds up an innermost region composed almost entirely of helium.
Since helium fusion requires significantly higher densities and temperatures, the helium core cannot burn its fuel and remains inert.
As the helium core contracts under gravity, the surrounding shell of hydrogen around the core becomes hotter and denser to ignite hydrogen burning at an increasingly rapid rate.
The resulting outward pressure causes the stellar envelope to expand and cool to make the star redder and more luminous as a transition into the RG branch (RGB) on the Hertzsprung–Russell (HR) diagram.
Driven by hydrogen burning, the helium-rich core grows in mass, so that its density and temperature increase and electron degeneracy develops.
Once the core temperature reaches approximately $10^8\,\mathrm{K}$, a thermonuclear runaway occurs through the triple-alpha process: ${}^4_2\mathrm{He} + {}^4_2\mathrm{He} \to {}^8_4\mathrm{Be} + \gamma$, followed by ${}^8_4\mathrm{Be} + {}^4_2\mathrm{He} \to {}^{12}_6\mathrm{C} + \gamma$.
This violent event, known as the helium flash, releases enough energy to lift the degeneracy of the core, halting the runaway and leading to a stable phase of core helium burning.
The star subsequently settles onto the horizontal branch (HB) in the HR diagram.
In standard stellar evolution, the core mass at the moment of helium ignition is a robust and well-predicted quantity.
Indeed, current stellar evolution models yield TRGB luminosities that closely match available observations \cite{Straniero:2020iyi}.
Hence, any deviations from this prediction, such as premature (or delayed) helium flash induced by an extra heating~\cite{Lopes:2021jcy,Dessert:2021wjx,Hong:2024ozz} (or cooling~\cite{Raffelt:1996wa,Capozzi:2020cbu,Li:2023vpv,Dolan:2023cjs}) mechanism, can leave observable imprints on stellar populations, especially in globular clusters.

\begin{figure}
    \centering
    \includegraphics[width=0.45\textwidth]{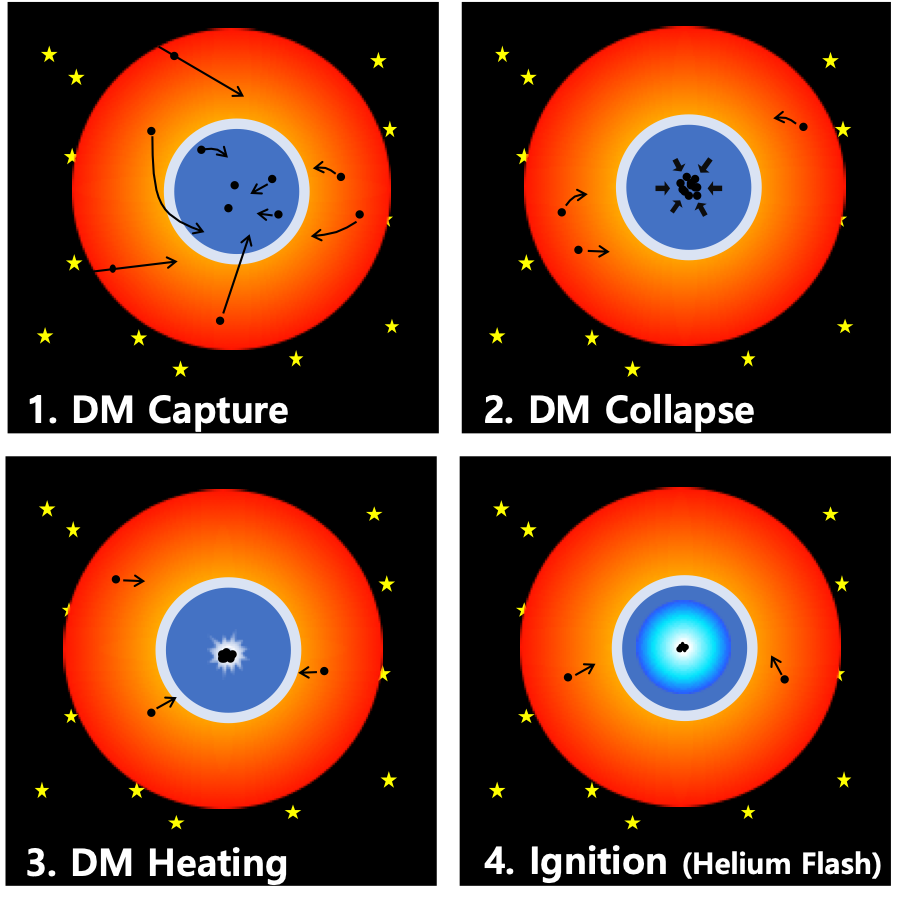}
    \caption{Schematic overview of the main phases of DM capture and its impact on RG evolution, proceeding in order as follows.
    \textit{Upper left}: DM particles are captured via elastic scatterings with nuclei (Sec.~\ref{sec:capture}).
    \textit{Upper right}: Captured DM concentrates near the stellar center through ingress (Sec.~\ref{sec:ingress}) and subsequently thermalizes (Sec.~\ref{sec:thermalization}). Once the DM density exceeds that of the ambient medium, the DM core gravitationally collapses (Sec.~\ref{sec:self-grav-collapse}).
    \textit{Lower left}: The collapsed DM core heats the surrounding medium. If the heating overcomes diffusive cooling, it can trigger helium ignition (Sec.~\ref{sec:TriggerMass}).
    \textit{Lower right}: Runaway helium burning occurs when DM heating satisfies the trigger mass condition (Sec.~\ref{sec:DMheatingVSdiffusion}).}
    \label{fig:scheme}
\end{figure}

In this work, we investigate the capture of DM particles by RG cores and their potential impact on stellar evolution, particularly on the timing of helium ignition.
The schematic overview is illustrated in Fig.~\ref{fig:scheme}.
DM particles lose kinetic energy through repeated elastic scatterings with nuclei in the stellar interior and become gravitationally bound once their velocity falls below the escape speed.
As they continue to lose energy, their orbits shrink deeper into the core; we refer to this process as ``ingress".
Eventually, the captured DM population thermalizes with the stellar medium, forming a compact DM sphere at the stellar center.
When the DM density within this sphere surpasses that of the surrounding stellar matter, its self-gravity dominates such that the gravitational collapse can occur.
This collapse deposits DM energy into the ambient material through elastic scattering and/or annihilation, depending on the underlying interaction model.
In particular, for HDM, since the pair-annihilation cross section is intrinsically small due to the unitarity bound, DM particles can accumulate to extremely high densities without being efficiently depleted by annihilation.
We highlight that this delayed annihilation can therefore provide a concentrated and intense source of heating at the stellar center.

If this exotic heating exceeds the rate of thermal diffusion, it can ignite a local helium flame that propagates outward through the core, driving a runaway helium ignition.
This scenario stands in sharp contrast to standard stellar evolution, where efficient neutrino cooling via the plasmon decay suppresses central ignition and instead causes helium burning to commence in an off-center shell of the degenerate core.
In our case, dynamical DM heating shifts the ignition site to the stellar center, because the DM heating can increase the temperature of the ignition site to a temperature higher than $ 10^8\,{\rm K}$ where the nuclear reaction rate dominates over the plasmon decay rate.
To quantify this process, we estimate the minimum amount of core material that must be heated to ignite helium and initiate this runaway fusion, namely the trigger mass for helium flash~\cite{1992ApJ...396..649T,Timmes_2000}.

A premature, DM-induced helium flash leads to a smaller helium core mass than in standard stellar evolution, resulting in a lower luminosity at the TRGB.
By requiring consistency with observed RG populations, we derive stringent constraints on the DM–nucleon scattering cross section.

The remainder of the paper is organized as follows.
In Sec.~\ref{sec:DM_capture}, we discuss the dynamics of DM particles captured by a star, including their concentration at the center through elastic scattering with the stellar medium.
Sec.~\ref{sec:DMheating} explores how a sufficiently dense and self-gravitating DM core can deposit energy into the surrounding material via elastic scatterings and/or annihilation and potentially trigger helium ignition in RG.
We provide an estimate of the trigger mass required for such a runaway process.
Finally, we conclude with a summary of our main findings.
In the Appendix, we provide useful formulas and expressions for our analysis.
For the sake of convenience, we adopt $\hbar=c=k_B=1$.









\section{Dark Matter Capture in Stars}
\label{sec:DM_capture}

In this section, we discuss the capture of DM particles ($\chi$) by a star and their subsequent concentration in the innermost region.
We divide the process into distinct phases: capture, ingress, thermalization, and gravitational collapse.
Throughout these stages, we assume that the dynamics is governed by elastic scatterings between DM and nuclei, as the DM number density is too low for the DM annihilation to significantly affect the system.
The conditions under which DM annihilation becomes relevant will be addressed in the next section.

DM particles originating from the galactic halos and located far away from a star can be gravitationally attracted by the star.
The differential incoming DM flux can be written as~\cite{Bramante:2017xlb,Acevedo:2019gre}
\bea
d \mathcal{F}_\chi = \pi \dfrac{f(u)}{u} \,du \, dJ^2\,,
\label{eq:flux_master}
\eea
where $J  = \omega(r) r \sin \theta$ is associated with the conserved angular momentum of a DM particle at radius $r$, $\omega(r) = \sqrt{v_e(r)^2 + u^2}$ is the speed of a DM particle at $r$ with the local escape velocity $v_e(r)$ and the initial velocity at infinity, and $\theta$ is the angle between the DM velocity and the radial vector. 
The function $f(u)$ denotes the DM velocity distribution at infinity, which we assume to follow a Maxwell-Boltzmann distribution throughout our analysis 
\bal
f(u) = 3 \sqrt{\dfrac{6}{\pi}} \dfrac{\rho_\chi u^2}{m_\chi\bar{v}^3}
\exp\left(-\dfrac{3 u^2}{2 \bar{v}^2}\right)\,,
\label{eq:MB_distribution}
\eal
where $m_\chi$ is the DM mass.
For numerical analysis, we take the local DM density of $\rho_\chi = 0.3\,{\rm GeV}/{\rm cm}^3$ and the averaged DM speed of $\bar{v} = 220\,{\rm km}/{\rm s}$.
Although the DM density in globular clusters is subject to significant uncertainties~\cite{Mashchenko:2004hj,Moore:2005jj,Saitoh:2005tt,Griffen:2009vg,Willman:2012uj,Tollerud:2010bj,Creasey:2018bgv,Madau:2019srr,Vitral:2021jvp,Garani:2023esk}, our capture results scale linearly with the assumed density, making our conclusions robust to such variations.

In the presence of non-gravitational interaction between DM and nuclei, DM particles can undergo elastic scattering with the stellar medium as they traverse the star.
If the kinetic energy of a DM particle significantly exceeds that of target nuclei (typically set by temperature), the fractional kinetic energy loss per collision with a target nucleus $T$ in the process $\chi + T \rightarrow \chi +T$ is given by 
\bal
\dfrac{\Delta K_\chi}{K_\chi} = 
-\dfrac{4 m_\chi m_T }{(m_\chi + m_T)^2} x \,,
\label{eq:energy_loss}
\eal
where $K_\chi$ denotes the kinetic energy of the incident DM,  $m_T$ is the mass of the target nucleus, and $x\in [0,1]$ is a random variable associated with the scattering angle in the center-of-mass frame.
For convenience, we define
\bal
\beta = \frac{4m_\chi m_T}{(m_\chi + m_T)^2} \, ,
\eal
which characterizes the efficiency of energy loss per collision.

\subsection{Capture}
\label{sec:capture}

A DM particle is captured by a star when it loses more energy through scattering than its initial kinetic energy at infinity, i.e., $m_\chi u^2/2$, during its passage through the stellar interior.

Given the fractional kinetic energy loss per collision described in \eqref{eq:energy_loss}, the probability that a DM particle with incoming kinetic energy $m_\chi \omega^2/2$ is captured after undergoing $N$ scatterings is given by
\begin{equation}
    g_N (\omega) = \left[\prod_{i = 1}^N\int_0^1 d x_i \, f (x_i)\right] \Theta \left(\dfrac{v_e^2}{\omega^2} - \prod_{i=1}^N (1-\beta x_i)\right) \,,
\label{eq:capture_prob}
\end{equation}
where $f(x_i)$ is the probability density function (PDF) associated with the angular variable $x_i$ for the $i$-th scattering. 
For simplicity, we assume that each PDF is uniformly distributed and that scatterings are independent.
In Appendix~\ref{sec:gN}, we derive an analytic approximation of $g_N(\omega)$ in the limits of $N \gg 1$ and $\beta \ll 1$ (equivalently $m_\chi \gg m_T$), where the linear expansion of the step function argument is allowed.

The trajectory of a DM particle within the star is determined by its conserved angular momentum $J$.
Since the escape velocity around the RG core ($\sim 0.01\,c$) exceeds the typical DM halo velocity ($\sim 10^{-3}c$), the DM path relies on $\theta$, defined as the relative angle between the DM velocity and the radial direction.
We define the core radius $R_c$ as the radius at which the escape velocity is $\sqrt{2/3}$ times the escape velocity at the center.
This $\sqrt{2/3}$ criterion corresponds to the escape velocity ratio for a uniformly dense sphere and aligns well with the straight-line approximation of particle trajectories.

Assuming a spherically symmetric stellar profile and adopting the straight-line approximation, the expected number of scatterings during the transit of DM through the star is given by the optical depth as follows
\bal
\tau(y) = 2 \int_0^{\Rc 
y} \,n_T(r) \sigma_{\chi T} \,d \ell\,,
\label{eq:tau_exact}
\eal
where $y \equiv \cos \theta$, $n_T$ is the number density of target nuclei, and $r = \sqrt{\Rc^2 + \ell^2 - 2 \Rc \ell \,y}$ is the radial distance along the path.
Here, $\sigma_{\chi T}$ denotes the elastic scattering cross section between DM and target nuclei.

For momentum transfer scales below the typical inverse nucleon size $(\sim 10\,{\rm fm} \sim 0.1\,{\rm MeV}^{-1})$, DM scatters coherently off the nucleons in a nucleus.
At higher momentum transfers, such a coherence is lost and the cross section becomes suppressed.
For heavy DM with $m_\chi \gg m_T$, this transition occurs at incident velocities around $v_\chi \sim 0.01c$.
We parametrize $\sigma_{\chi T}$ in terms of the DM-nucleon elastic scattering cross section $\sigma_{\chi N}$, which is assumed to be spin-independent, isospin symmetric, and velocity independent, as follows
\bea
\sigma_{\chi T} \simeq \sigma_{\chi N} A^2\left(\frac{\mu_{\chi T}}{\mu_{\chi N}}\right)^2 F^2\left(E_R\right)\,,
\eea
where $A$ denotes the atomic mass number of the target nucleus, $\mu_{\chi T}$ ($\approx A \mu_{\chi N}$ for $m_\chi \gg m_T$) is the DM-nucleus reduced mass, and $F(E_R)$ is the nuclear form factor correction in terms of the recoil energy $E_R$~\cite{Helm:1956zz}.
This form factor is of order unity for $E_R \lesssim 1\,{\rm MeV}$ and becomes suppressed for higher recoil energies.

Considering that each infinitesimal segment along the path contributes equally to the total optical depth, the probability for the $N$ number of scatterings with the given optical depth in \eqref{eq:tau_exact} follows the Poisson distribution function
\bal
p_N(y)
= \dfrac{\tau(y)^N}{N!} e^{-\tau(y)}\,.
\label{eq:poisson}
\eal

The DM capture rate can be expressed as~\cite{Bramante:2017xlb}
\begin{equation}
\begin{split}
\dot{{\cal C}} = & \int d \mathcal{F}_\chi \bigg(\sum_N g_N p_N\bigg)\Theta\bigg(\omega \Rc - J\bigg) \\
= & \dfrac{3 \pi \Rc^2}{\bar{v}^3}
\sqrt{\dfrac{6}{\pi}}
\dfrac{\rho_\chi}{m_\chi}
\int_0^\infty du \,u \,\omega^2\,\exp\left(-\dfrac{3 u^2}{2\bar{v}^2}\right)  \\
& \times \sum_{N=1}^{\infty} g_N \left(2\int_0^1 dy\,y \,p_N\right)\,.
\end{split}
\label{eq:capture_rate}
\end{equation}
Here, we take the escape velocity at the core radius, $v_e = v_e(R_c)$, based on the $\sqrt{2/3}$ criterion discussed above, which offers a conservative estimate under the straight-line approximation.
In the remainder of this work, we adopt the escape velocity defined at the core radius.

Assuming a homogeneous density $n_T$,
the capture rate scales linearly with $\sigma_{\chi T}$ in the small cross section regime, $\sigma_{\chi T} \ll (1/\beta n_T\Rc)(\bar{v}^2/v_e^2)$.
This behavior arises because DM particles with low initial velocities, $u^2 \lesssim (2v_e^2/3)(\beta n_T\Rc)\sigma_{\chi T} \ll \bar{v}^2$, are more likely to be captured through a typical number of scatterings (i.e., the total optical depth $\tau$).
As a result, the capture probability becomes proportional to the optical depth and thus scales linearly with the cross section.

Conversely, when the cross section exceeds a critical value, $\sigma_{\chi T} \gtrsim (1/\beta n_T\Rc)(\bar{v}^2/v_e^2)$, most incoming DM particles undergo sufficient scatterings to be captured.
In this saturation regime, the capture rate reaches the geometric limit and becomes effectively independent of $\sigma_{\chi T}$. 

We find that the analytic approximation for the capture rate provides a good fit to the full numerical results 
\bea
\dot{\mathcal{C}} & \simeq & \dot{\mathcal{C}}_{\rm geo} \times  \left\{
\begin{tabular}{cc}
    $\left(\dfrac{\sigma_{\chi T}}{\sigma_{\rm thr}}\right)$ & for~$\sigma_{\chi T} \ll \sigma_{\rm thr}$ \vspace{0.15cm}  \\
    $1$ & for~$\sigma_{\chi T} > \sigma_{\rm thr}$
\end{tabular}
\right.\,,
\label{eq:AnalyticCapture}
\eea
where the capture rate in the geometric limit is given by
\begin{align}
    \dot{\mathcal{C}}_{\rm geo} & \equiv \pi R_c^2 \frac{\rho_\chi}{m_\chi}\bar{v} \sqrt{\frac{6}{\pi}} \left(\frac
{v_e^2}{\bar{v}^2}\right) \, ,
\label{eq:geocapturerate}
\end{align}
and the threshold cross section is defined as
\begin{align}
    \sigma_{\rm thr} &\equiv \frac{2}{\beta n_T\Rc}\left(\frac{\bar{v}^2}{v_e^2}\right) \, .
    \label{eq:thresholdCrossSection}
\end{align}



In realistic stellar environments, the interior medium typically consists of multiple nuclear species.
This multiple species case can be appropriately addressed by generalizing the quantities for the fractional kinetic energy loss $\beta$ and the optical depth $\tau$ as follows:
\bea
\beta & = & \frac{1}{\sum_i n_i \sigma_{\chi i}} \sum_{i} n_i\sigma_{\chi i} \dfrac{4 m_\chi m_i}{(m_\chi + m_i)^2}\,,
\label{eq:beta_multiple_target} \\
\tau(y) & = & \sum_{i}2\int_0^{\Rc 
y} n_i(r) \, \sigma_{\chi i} \,d\ell\,,
\label{eq:tau_multiple_target}
\eea
where $n_i$ is the number density of nuclei of species $i$, and $\sigma_{\chi i}$ is the corresponding DM–nucleus scattering cross section. 

Finally, the total number of DM particles captured over the stellar lifetime can be obtained by integrating the time-dependent capture rate
\bal
N_\chi (t) = \int \dot{{\cal C}} (t^\prime)\, dt^\prime\,.
\label{eq:Ncapture}
\eal


\subsection{Ingress}
\label{sec:ingress}

Captured DM particles are gravitationally bound to the star, but typically follow elliptic orbits that remain mostly outside the stellar radius.
Assuming the usual case of high orbital eccentricity, the total energy of a DM particle is approximately $E_\chi = - G \Mc m_\chi/b$ where $\Mc$ is the stellar core mass, and  $b$ denotes the `apastron' distance (i.e., the farthest point from the stellar center).
While a DM particle is inside the star ($r < \Rc$), its kinetic energy is given by
\bal
K_\chi(r) = E_\chi - V_\chi(r) = E_\chi + \dfrac{G \Mc m_\chi}{2 \Rc^3} \left(3 \Rc^2 - r^2\right)\,\,.
\eal
Considering the averaged fractional energy loss for each collision given by $ \beta /2$ and the orbital period of $2 \pi (b/2)^{3/2}/\sqrt{G \Mc}$, the DM energy loss rate can be estimated as
\bea
\frac{d E_\chi}{dt} = -\frac{\sqrt{2G \Mc}}{ \pi b^{3/2}}  \beta \int_0^{\Rc} dr\, K_\chi(r) n_{T}(r) \sigma_{\chi T} \,.
\label{eq:dE_dt_gen}
\eea
Here, we approximate the evolution using a single representative DM trajectory that intersects the stellar radius.
Using the relation $dE_\chi = (G \Mc m_\chi/b^2) db$, \eqref{eq:dE_dt_gen} can be rewritten in terms of $b$ as
\bea
\dfrac{db}{dt} = -\dfrac{\sqrt{2b}}{\pi \sqrt{G \Mc} m_\chi} \beta
\int_0^{\Rc} dr\, K(r)  n_{T}(r) \sigma_{\chi T}\,.
\label{eq:db_dt_gen}
\eea
Under the assumption of a uniform stellar density, \eqref{eq:db_dt_gen} simplifies to
\begin{equation}
\begin{split}
\dfrac{db}{dt} &= -\dfrac{\sqrt{2b}}{\pi \sqrt{G \Mc} m_{\chi}}
 \beta \dfrac{\tau_c}{2 \Rc} \int_0^{\Rc} dr\, K(r)\,\\
& = -\dfrac{\sqrt{G M_c}  \beta \tau_c}{\sqrt{2}\pi \Rc} \sqrt{b} \left(\dfrac{4}{3} - \dfrac{\Rc}{b}\right)
\end{split}
\label{eq:db_dt_uni}
\end{equation}
where $\tau_c \equiv \tau(1) \simeq 2   n_{T} \sigma_{\chi T} \Rc$
represents the expected number of scatterings during a single crossing through the stellar core.


The average apastron distance of captured DMs, $\left<b\right>$, relies on the DM-nucleus elastic scattering cross section, $\sigma_{\chi i}$.
Given that each scattering reduces the energy by an average fraction $\beta/2$, and the total number of scatterings for a crossing over the stellar center is $\tau_c$, the average energy of captured DMs is straightforwardly evaluated as $\left<E_\chi\right> \approx - m_\chi \beta\tau_c v_e^2/4$.
This yields an estimate for the average apastron distance as
\bea
\left<b\right> \simeq \frac{2\Rc}{\beta \tau_c}\,.
\eea

The ingress phase ends when the DM orbit lies within the stellar core.
The time scale for this inward migration can be estimated from \eqref{eq:db_dt_uni} as~\footnote{This expression exhibits a different parametric dependence, particularly on the elastic scattering cross section $\sigma_{\chi i}$, compared to Ref.~\cite{Acevedo:2019gre}, due to a different assumption on the initial apastron distance.} 
\begin{equation}
\begin{split}
t_{\rm ing} 
& \simeq \frac{3\pi\Rc^{\frac{3}{2}}}{\sqrt{GM_c}\beta^{\frac{3}{2}}\tau_c^{\frac{3}{2}}} \\
& = 8.6\,{\rm Gyr} \,\rho_6^{-\frac{3}{2}}\left(\frac{M_c}{0.1 M_\odot}\right)^{-\frac{1}{2}} A^{-6}\\
& \quad \times \left(\frac{m_\chi}{10^{10}\,\rm GeV}\right)^
{{\color{blue}}\frac{3}{2}} \left(\frac{\sigma_{\chi N}}{10^{-40}\,\rm cm^2}\right)^{-\frac{3}{2}}  \,,
\end{split}
\label{eq:tingress_uni_exact}
\end{equation}
assuming $\langle b \rangle \gg \Rc$.
In the final expression, we approximate $\sigma_{\chi T} \approx A^4\sigma_{\chi N}$ and $m_T \simeq A m_N$, with the atomic number $A$.
For notational convenience, we define the normalized variables
\bea
T_8 \equiv \frac{T_c}{10^8\,{\rm K}}\,, \quad \rho_6 \equiv \frac{\rho_c}{10^6\,\rm g/cm^3} \,,
\eea
for the core temperature $T_c$ and the core density $\rho_c$.

\subsection{Thermalization}
\label{sec:thermalization}

After the ingress phase, DM particles become fully confined within the stellar interior and undergo repeated collisions with the surrounding medium.
The average fractional energy loss per elastic scattering with stellar nuclei is given by 
\bea
\label{eq-energyloss}
\left< \Delta K_\chi \right>
= \dfrac{\beta}{2} \left(K_\chi - \frac{3}{2}T_c\right) \,
\eea
with the core temperature $T_c$.
This expression shows that DM particles continue to lose energy until their kinetic energy becomes comparable to the thermal energy of the stellar medium, i.e., $K_\chi \sim \tfrac{3}{2} T_c$.
Notably, in the high-energy limit where $K_\chi \gg T_c$ (e.g., the capture phase), this formula recovers the earlier description in \eqref{eq:energy_loss}.

The evolution of the DM kinetic energy is governed by the Boltzmann equation as follows
\bea
\dfrac{d K_\chi}{d t} = -  n_T \sigma_{\chi T} v_{\rm rel} \frac{\beta}{2} \left(K_\chi-\frac{3}{2} T_c \right) \,,
\eea
where the average relative velocity between DM and nuclei is approximately $v_{\rm rel} \approx \sqrt{2K_\chi/m_\chi + 3T_c/m_T}$.
This equation indicates that the thermalization timescale $t_{\rm th}$ is sensitive to $K_\chi$ in the low-energy regime, particularly when the relative velocity is dominated by the thermal motion of nuclei, i.e., $v_{\rm rel} \approx \sqrt{3T_c/m_T}$.
This results in
\begin{equation}
    \begin{split}
        t_{\rm th} & \simeq \dfrac{2}{n_T \sigma_{\chi T} \beta} \sqrt{\frac{m_T}{3T_c}} \log\left[\frac{3m_\chi}{m_T}\right] \\
        & = 1.8 \times 10^{-5}\,{\rm Gyr} \,\rho_6^{-1}\,T_8^{-\frac{1}{2}}A^{-\frac{7}{2}}\\
        & \quad \times \left(\frac{\sigma_{\chi N}}{10^{-40}\,{\rm cm}^2}\right)^{-1}\left(\frac{m_\chi}{10^{10}\,{\rm GeV}}\right) \log\left[\frac{3m_\chi}{m_T}\right] \, .
    \end{split}
    \label{eq:thermaltimescale}
\end{equation}
Note that, formally, it takes an infinite time for the kinetic energy to reach exactly $ K_\chi=3T_c/2 $ through scatterings.
For practicality, we define thermalization as complete when the kinetic energy drops to $K_\chi = 2 T_c$, which sufficiently ensures equilibrium with the stellar environment.

\subsection{Gravitational Collapse}
\label{sec:self-grav-collapse}

Once thermalized, DM particles concentrate near the stellar core and settle into a compact configuration at the center.
In the early stages, when the DM density remains low, the gravitational dynamics of the thermalized DM population are dominated by the background stellar potential.
In this regime, considering the typical DM kinetic energy of $3T_c/2$, the virial theorem implies that the characteristic size of the thermalized DM sphere is
\begin{equation}
\begin{split}
    r_{\rm th} & =  \sqrt{\dfrac{9 T_c}{4 \pi G \rho_{\rm c} m_{\chi}}} \\
    & = 2.8\times 10^3\,{\rm cm}\,\left(\dfrac{m_\chi}{10^{10}\,\rm GeV}\right)^{-\frac{1}{2}} T_8^{\frac{1}{2}}\rho_6^{-\frac{1}{2}}\,.
\end{split}
\label{eq:rth}
\end{equation}

The total number of DM particles within the thermalized sphere continues to grow through continuous accumulation via the capture process, as described in Sec~\ref{sec:DM_capture}.
Once the enclosed DM mass exceeds that of the ambient stellar material within the same radius, the configuration becomes self-gravitating.
The threshold mass for the onset of self-gravitation is approximately
\begin{equation}
\begin{split}
    M_{\rm sg} & =  \frac{4\pi \rho_{\rm c} r_{\rm th}^3}{3} \\
     & =  0.9 \times 10^{17} {\rm g} \left(\dfrac{m_\chi}{10^{10}\,\rm GeV}\right)^{-\frac{3}{2}} T_8^{\frac{3}{2}}\rho_6^{-\frac{1}{2}}\,.
\end{split}
\label{eq:Mth}
\end{equation}

The timescale for reaching this self-gravitating phase is minimized when the DM capture rate saturates to the geometric limit as described in \eqref{eq:geocapturerate}.
Under this condition, the time required to accumulate the critical number of DM particles is given by
\begin{equation}
    \begin{split}
       \frac{M_{\rm sg}/m_\chi}{\dot{{\cal C}}_{\rm geo}}   & \approx 4.2 \times 10^{-3}\,{\rm Gyr} \, \rho_6^{-\frac{1}{6}} T_8^{\frac{3}{2}}\left(\frac{M_c}{0.1 M_\odot}\right)^{-\frac{4}{3}} \\
       \times & \left(\dfrac{m_\chi}{10^{10}\,\rm GeV}\right)^{-\frac{3}{2}}\left(\dfrac{\rho_\chi}{0.3\,{\rm GeV}/{\rm cm}^3}\right)^{-1}\left(\dfrac{\bar{v}}{10^{-3}c}\right)\,.
    \end{split}
\end{equation}
Hence, for DM masses below $10^{8}\,{\rm GeV}$, the capture rate during the RG phase is insufficient to achieve the collapse condition.

Once the total mass of captured DM particles exceeds the threshold $M_{\rm sg}$, the DM sphere becomes self-gravitating and dynamically unstable, which initiates gravitational collapses. 
The characteristic free-fall timescale is given by
\bea
t_{\rm ff} = \sqrt{\dfrac{1}{4 \pi G \rho}} = 1.09 \,{\rm sec} \, \rho_6^{-\frac{1}{2}}\,.
\label{eq:tff}
\eea
This timescale is typically much shorter than any relevant timescale associated with pressure-supporting interactions, which ensures that the DM configuration is indeed gravitationally unstable once the collapse condition is met.
However, the collapse does not result in an immediate black hole formation; instead, the DM particles rapidly virialize on a timescale comparable to $t_{\rm ff}$.

During this phase, DM particles gain increasingly high velocities, elevating their kinetic energy above the ambient core temperature.
Subsequent collisions with stellar nuclei cause the DM core to dissipate energy and shrink further in size.
This contraction continues until the DM velocity approaches $\sim 0.02\,c$, beyond which the nuclear form factor becomes strongly suppressed, significantly reducing the energy loss efficiency.
Beyond this velocity, scattering could potentially disrupt nuclei, but this is beyond the scope of this paper.
For analysis, we conservatively terminate the collapse process when the typical DM velocity reaches $v_{\rm cut} = 0.02\,c$ and do not consider black hole formation (and following heating via evaporation).
The physical consequences of such a collapse, particularly in terms of energy deposition and heating of the surrounding stellar material, will be explored in the next section.

\subsection{Numerics}
\label{sec:numerics}

By means of the FuNS (Full Network Stellar Evolution) code, we perform numerical analyses based on low-mass stellar models with initial stellar parameters (mass and composition) relevant to RGB stars in galactic globular clusters.
A detailed description of the code and its setup is reported in Ref.~\cite{Straniero:2020iyi}. 
In particular, we adopt an initial stellar mass of $0.83\,M_\odot$, metallicity of $Z = 0.001$, and initial helium abundance $Y=0.25$. 
The initially homogeneous stellar structure then evolves through the pre-main sequence, main sequence, and RGB phases, up to the helium flash at the TRGB, where the model age is 12.2 Gyr, close to the average globular cluster age (11-13 Gyr).

Conventionally, the helium flash is assumed to begin when the luminosity generated by the activation of the triple-$\alpha$ nuclear reactions is $10^2$ times the stellar luminosity.
For the adopted model, this corresponds to a TRGB bolometric magnitude of $M_{bol}=4.75-2.5\log L/L_\odot=-3.64$ that aligns with the measured values for a globular cluster with a low-metallicity of about Z=0.001.
For instance, the well-studied globular cluster NGC 5904 (M5) in the northern hemisphere, which has nearly the same metallicity as our stellar models, exhibits a TRGB bolometric magnitude of $M_{\rm bol} = -3.63 \pm 0.26$ (see Table 6 of Ref.~\cite{Straniero:2020iyi}). 
The consistency between theory and observation implies that any non-standard physics can be accommodated only if its effect on the TRGB luminosity remains within observational uncertainties.

Accordingly, if DM capture process (including ingress, thermalization, and self-gravitating collapse) takes place within the RGB timescale as described in previous subsections, we assess whether the helium flash is triggered earlier, at a time $t_{\rm end}$, when the stellar luminosity falls short of the standard TRGB value by $\log_{10}[|\Delta L|/L_\odot] = 0.13$, corresponding to a bolometric magnitude offset of $\Delta M_{\rm bol} = 0.325$.
This threshold closely matches the conservative $95\%$ confidence level adopted in Ref.~\cite{Straniero:2020iyi}, ensuring that our predictions remain consistent with observations.
We further assume that the stellar structure is unaffected by DM capture prior to $t_{\rm end}$, so the only modification is an earlier helium flash induced by heating from the DM core collapse.

\begin{figure}
    \centering
    \includegraphics[width=0.45\textwidth]{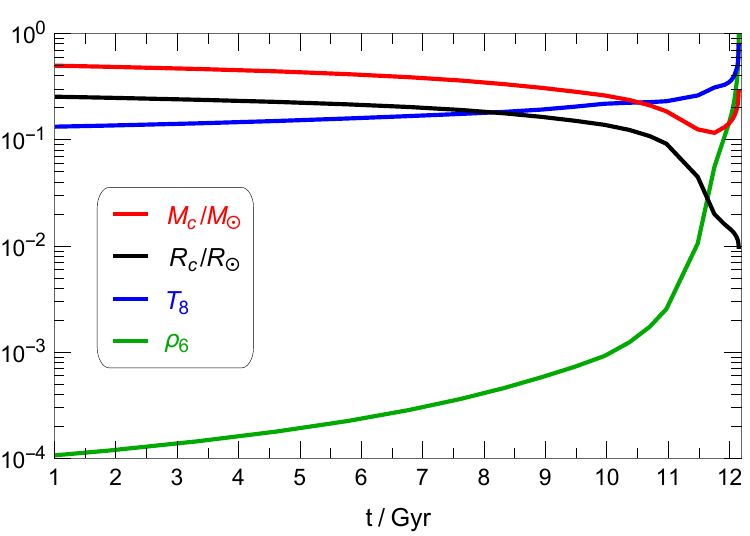}
    \caption{Evolution of the core mass (red), core radius (black), central temperature (blue), and central density (green). See the text for our operational definition of the `core', based on the escape-velocity criterion relevant for DM capture.}
    \label{fig:Mcore_Rcore_T0_vs_time}
\end{figure}

Fig.~\ref{fig:Mcore_Rcore_T0_vs_time} illustrates the key profile parameters of the core radius $\Rc$ (black), the core mass $M_{\rm c}$ (red), the core temperature $T_c$ (blue), and the core density $\rho_c$ (green) from $1\,{\rm Gyr}$ up to $t_{\rm end}$.
As discussed above, we define the core boundary in terms of the escape velocity profile, taking $\Rc$ to be the radius at which $v_{\rm esc} (\Rc)=\sqrt{2/3}v_{\rm esc} (0)$.
This definition captures the approximately homogeneous central region, where the density profile is nearly flat.
The resulting homogeneity supports the straight-line approximation for incoming DM trajectories within the region that dominates the scattering probability.
This restriction is conservative, as it neglects potential scattering sites at larger radii and because realistic trajectories are generally more centrally focused than straight lines.
Moreover, we verify that the optical depth controlling DM interactions and capture is dominated by this core region, so that our semi-analytic estimates remain consistent in the context of the full stellar model.

\begin{figure}
    \centering
    \includegraphics[width=0.45\textwidth]{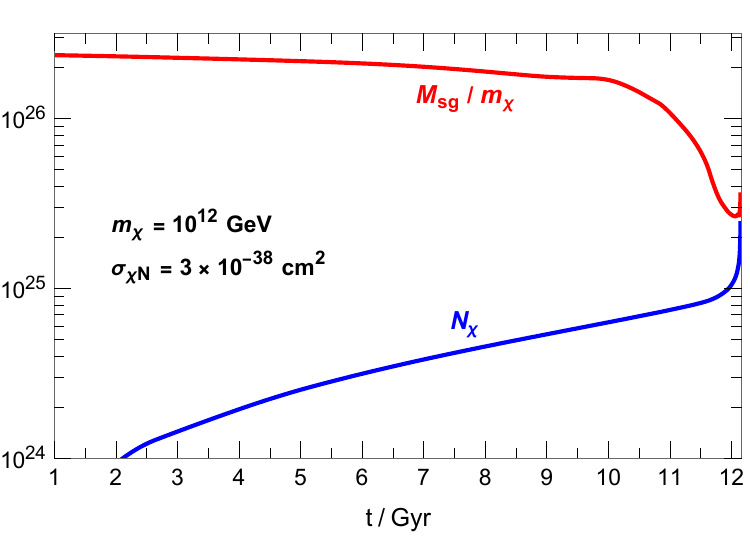}
    \caption{Evolution of the total number of accumulated DM particles $N_{\chi}$ (blue) and the threshold number for self-gravitation $M_{\rm sg}/m_\chi$ (red), with the DM mass of $m_{\chi} = 10^{12}\,{\rm GeV}$ and the DM-nucleon cross section of $\sigma_{\chi {\rm N}} = 3\times 10^{-38}\,{\rm cm}^2$.
    }
    \label{fig:DM_number_vs_time}
\end{figure}

In Fig.~\ref{fig:DM_number_vs_time}, we show the total number of captured DM particles (blue curve) as a function of time, for a benchmark mass of $m_\chi = 10^{12}\,\mathrm{GeV}$ and a DM–nucleon cross section of $\sigma_{\chi N} = 3\times 10^{-38}\,\mathrm{cm}^2$.
A sizable number of DM particles is captured during the RG phase, despite its relatively short duration of approximately $1\,\mathrm{Gyr}$.
This behavior is consistent with the scaling in \eqref{eq:AnalyticCapture}.
In the weak interaction regime, where the capture rate scales linearly with the elastic scattering cross section, we find $\dot{\mathcal{C}}\propto M_c^3/R_c^2$.
Using characteristic core parameters for the main sequence ($M_c \simeq 0.5 M_\odot$, $R_c \simeq 0.2 R_\odot$) and for the RG phase ($M_c \simeq 0.2 M_\odot$, $R_c \simeq 0.01 R_\odot$) as shown in Fig.~\ref{fig:Mcore_Rcore_T0_vs_time}, the capture efficiency during the RG phase is enhanced by a factor of roughly 30 compared to the main sequence.
This enhancement compensates for the shorter duration, which results in a total number of captured particles during the RG phase that is comparable to the main sequence contribution.
For comparison, we also plot the evolution of the critical number of DM particles required for the onset of self-gravitation ($N_{\rm sg}$, red), which signals the transition to a gravitationally unstable DM core and potential collapse.
This comparison provides a useful check on whether the captured DM population can trigger sufficient localized heating to ignite helium burning.
As the plot indicates, the sharply increased capture rate during the RG phase makes the final snapshot particularly relevant for determining whether the captured DM population reaches the collapse threshold, as we will examine in detail in the next section.

\section{Dark Matter-Induced Helium Ignition in Red Giants}
\label{sec:DMheating}

We now turn to the potential consequences of DM collapse, most notably the premature ignition of helium in the RG core, and discuss possible observational implications.
If the heating induced by the collapsing DM core is efficient enough, it can raise the temperature of a portion of the surrounding stellar material to the point where the nuclear reaction rate for helium burning, specifically the triple-$\alpha$ process,
becomes large enough to sustain itself.
This can trigger a thermal runaway, ultimately igniting the entire helium core before the standard temperature and pressure conditions for helium flash are met.

Although the helium flash itself is not directly observable since most of the released energy is absorbed in lifting the electron degeneracy of the core, its premature onset can leave observable imprints.
One such consequence is a decrease in the luminosity at the TRGB.
In standard stellar evolution, initially low-mass stars (with $M \lesssim 2 M_\odot$) undergo helium ignition when the core reaches a temperature of $T_{\rm c} \sim 10^8\,{\rm K}$ and a density of $\rho_{\rm c} \sim 10^6\,{\rm g}/{\rm cm}^3$.
However, if the helium flash is triggered earlier (e.g., at a lower core mass due to DM-induced heating), the correlated tip luminosity of RG would be reduced.
Such variations can provide a novel avenue to constrain DM properties.
Moreover, since the TRGB has been proposed as a type of standard candle, premature helium ignition may have implications for the reliability of this distance indicator.

The condition for helium ignition is determined by the balance between nuclear heating and conductive or radiative cooling within a localized region of the core.
This leads to the concept of a ``trigger mass" $M_{\rm tr}$, defined as the minimal mass of core material that must reach a critical temperature for thermal runaway to occur.
As for the contribution from plasmon neutrino cooling~\cite{1996ApJS..102..411I}, 
we have confirmed that it is important only around TRGB, while it is subdominant when DM heats up the core material above $ 10^8 $K. Therefore, we can safely neglect this effect in our discussion.
In the following, we examine whether a collapsing DM core can deposit sufficient energy (through elastic scatterings or annihilation) to heat a region of mass $M_{\rm tr}$ above the ignition threshold.

\subsection{Trigger mass for helium ignition}
\label{sec:TriggerMass}

We begin by considering a localized, overheated region that forms within the helium core.
The excess heat from this region begins to diffuse outward via conduction or radiative diffusion, effectively cooling the region.
Simultaneously, the elevated temperature enhances helium burning through the triple-$\alpha$ process, which generates nuclear energy and may further raise the temperature.
Whether the thermal disturbance dissipates or triggers helium ignition depends on the competition between heat loss and nuclear energy generation~\cite{1992ApJ...396..649T}.

This competition is characterized by two timescales: the diffusion timescale $t_{\rm diff}$ and the nuclear burning timescale $t_{\rm burn}$.
If $t_{\rm diff} \gg t_{\rm burn}$, nuclear energy accumulates faster than it is lost, which results in a thermonuclear runaway that ignites the entire helium core.
Conversely, if $t_{\rm diff} \lesssim t_{\rm burn}$, the excess heat is efficiently conducted away before runaway conditions are achieved.

For a heated region of characteristic size $\delta$ with temperature $T_b$, the diffusion timescale is approximated as
\bea
    t_{\rm diff} \sim \frac{\delta^2}{D} \,,
\eea
where $D$ is the thermal diffusivity of the medium.
The nuclear burning timescale reads
\bea
t_{\rm burn} \sim \frac{\epsilon_{\rm th}}{\dot{S}} \,,
\eea
where $\epsilon_{\rm th}$ denotes the excess thermal energy per unit mass, and $\dot{S}$ is the local nuclear energy generation rate.
Equating these two timescales, $t_{\rm diff} \sim t_{\rm burn}$, yields the minimum size of $\delta$ necessary for runaway helium ignition.
This defines the trigger mass~\cite{1992ApJ...396..649T,Timmes_2000}
\begin{align}\label{eq-Mtr}
    M_{\rm tr} = \frac{4 \pi}{3} \delta^3 \rho_{\rm c} \sim \frac{4 \pi}{3} \left( \frac{\lambda_{\rm eff}  v_c \epsilon_{\rm th}}{\dot{S}} \right)^{\frac{3}{2}} \rho_{\rm c} \, ,
\end{align}
where we use $D\sim \lambda_{\rm eff} v_c $, with the effective mean free path $\lambda_{\rm eff}$ and the thermal velocity of the conducting particles $v_c$.
The total energy input required to heat this region is roughly given by $M_{\rm tr} \epsilon_{\rm th}$.
This order-of-magnitude estimate follows the treatment in Ref.~\cite{1992ApJ...396..649T}, and agrees well with more detailed numerical results.

The excess thermal energy per unit mass can be expressed as
\begin{equation}
    \epsilon_{\rm th} = \int_{T_c}^{T_b} c_m \, dT \,,
\end{equation}
where $c_m$ is the specific heat per unit mass, defined as $c_m \equiv  \left(\partial \epsilon_{\rm f}/\partial T \right) / \rho_{\rm c}$, with the thermal free energy density $\epsilon_{\rm f}$.
For a non-relativistic, fully ionized medium, the heat capacity is controlled by contributions from electrons and helium nuclei.
In the helium-rich medium, we exploit the following analytic approximations
\bea
\left[c_m\right]_{e^-} & = & \left\{ 
\begin{array}{cl}
    \dfrac{m_e p_{\rm F} T}{3 \rho_{\rm c}} & 
    \textrm{for}\ T<T_{\rm deg} \vspace{0.2cm}  \\ 
    \dfrac{3}{m_{\rm He}} & \textrm{for} \  T \geq T_{\rm deg}
\end{array}  
\right. \,, \\
\left[c_m\right]_{\rm He} & = & \frac{3}{2 m_{\rm He}} \,,
\eea
where $p_F = (3\pi^2 n_e)^{1/3}$ is the electron Fermi momentum, and $T_{\rm deg} = (p_F^2/2m_e)(3/\pi^2)$.
These approximations are valid for temperatures below $10^9\,{\rm K}$, beyond which electrons become relativistic and a contribution from photons is non-negligible.
In this work, we focus on the non-relativistic regime.

Next, we estimate the effective heat conductivity in the RG helium core, which determines how DM-induced heating competes with conductive cooling.
Since thermal conduction receives contributions from both radiative (via photon) and conductive (via electron) diffusions~\cite{1980SvA....24..303Y,Lampe:1968zz,1980SvA....24..126U}, the effective mean free path for conducting particles in the material reads
\bea
    \lambda_{\rm eff}= \lambda_e + \lambda_r \,,
\eea
where $ \lambda_e $ and $ \lambda_r $ are the mean free paths for electron and photon diffusion, respectively.
A larger $\lambda_{\rm eff}$ implies more efficient conduction, as energy is transported over longer distances.
We provide the detailed derivation of the mean free paths in Appendix~\ref{app:mfp}.

The energy generation rate from the triple-$\alpha $ process reads~\cite{1983ARA&A..21..165H}
\begin{equation}
\begin{split}
    \dot{S} &= \frac{Q_{3\alpha} r_{3\alpha}}{\rho_{\rm core}}  \\ 
    &\simeq 3.2\times 10^{26} \,\frac{\rm GeV}{{\rm s}\cdot{\rm g}} \,\rho_6^2 \, T_{b8}^{-3}  \exp \left[-\frac{44}{T_{b8}}\right] \,,
\end{split}
\end{equation}
where $Q_{3\alpha} = 3m_{^4{\rm He}} - m_{^{12}{\rm C}} = 7.275\,\text{MeV}$ is the net energy release per reaction, $r_{3\alpha}$ is the resonant reaction rate of the triple-$\alpha$ process, and we define $T_{b8} \equiv T_b/10^8{\rm K}$.

\begin{figure}
    \centering
    \includegraphics[width = 0.45\textwidth]{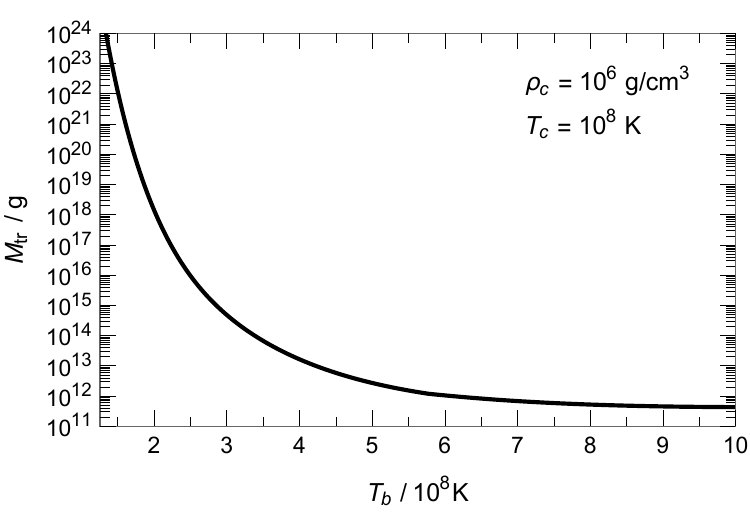}
    \caption{
    The trigger mass for the helium ignition, $M_{\rm tr}$, as a function of burning temperature $ T_b $, with $\rho_c = 10^6\,{\rm g}/{\rm cm}^3$ and $T_c = 10^8\,{\rm K}$ as the typical conditions of the RG core.
    }
    \label{fig:TriggerM}
\end{figure}

\begin{figure}
    \centering
    \includegraphics[width = 0.45\textwidth]{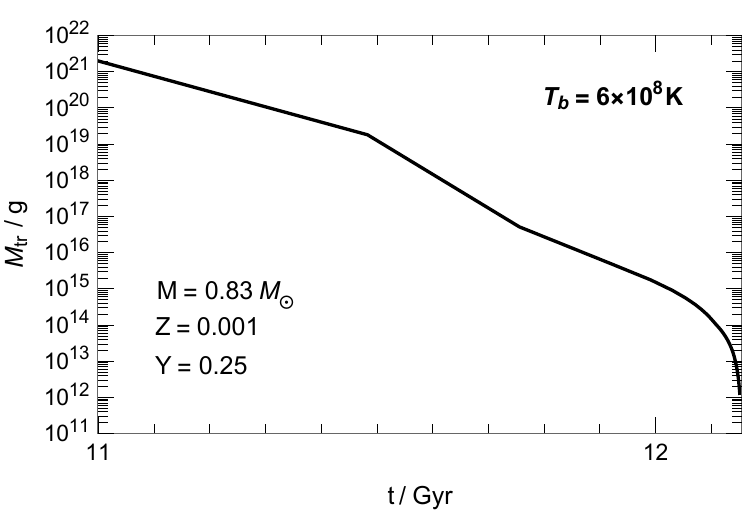}
    \caption{
    Evolution of the trigger mass for the helium ignition, $M_{\rm tr}$, based on the stellar evolution simulation.}
    \label{fig:TriggerMintime}
\end{figure}
 
Substituting the above expressions for $\epsilon_{\rm th}$, $\dot{S}$, and $\lambda_{\rm eff}$ into \eqref{eq-Mtr}, we obtain the trigger mass required to initiate helium ignition.
Fig.~\ref{fig:TriggerM} shows $M_{\rm tr}$ as a function of the burning temperature $T_b$, assuming the core temperature $T_8 =1$ and density $\rho_6 = 1$.
We confirm that our results are in good order-of-magnitude agreement with previous estimations of the trigger mass in Ref~\cite{Timmes_2000}.
We also demonstrate the time evolution of the trigger mass, based on our stellar evolution simulation with the benchmark burning temperature of $T_{b8} = 6$, the result of which is shown in Fig.~\ref{fig:TriggerMintime}.

In summary,  if DM-induced heating raises the temperature within a region of mass greater than $M_{\rm tr}$, such that the point lies above the corresponding curve in Fig.~\ref{fig:TriggerMintime} at a respective time, the released nuclear energy from helium burning can sustain the temperature and ignite the surrounding material.
This localized ignition then initiates a thermal runaway that propagates through the helium core, eventually triggering the helium flash.


\subsection{DM heating and diffusion}
\label{sec:DMheatingVSdiffusion}

Once DM particles in a RG form a self-gravitating core, they undergo gravitational free-fall and gradually lose energy through interactions with ambient helium nuclei.
This leads to a contraction of the DM core.
There are two key processes contributing to this contraction: elastic scatterings with nuclei, as discussed earlier, and DM self-annihilation, which becomes increasingly important as the DM number density rises during collapse.

For efficient heating of the stellar medium, we require that annihilation becomes significant only after the DM core collapses to sufficiently high densities.
If annihilation occurs too early, it can deplete the DM population before the core reaches the conditions necessary for efficient energy deposition.
A conservative criterion is that the DM annihilation timescale prior to collapse exceeds the typical RG lifetime $\sim 1 {\rm Gyr}$.

We denote the annihilation cross section by $\sigma_{\rm ann}$. 
In the context of the unitarity limit for the DM annihilation, we parameterize $\sigma_{\rm ann} v_\chi$ in terms of the DM mass as
\bea \label{eq:ann_cross}
\sigma_{\rm ann}v_\chi \equiv \frac{\alpha_\chi^2}{m_\chi^2}\,,
\eea
where $\alpha_\chi$ accounts for an effective coupling between DM and SM particles.
Here, we assume that the $s$-wave contribution is not suppressed, so that $\sigma_{\rm ann}v_\chi$ remains constant.

The annihilation timescale at the onset of collapse, where DMs are thermalized as described above, is given by
\bea
\left[ t_{\rm ann} \right]_{\rm col} = \frac{1}{n_\chi^{\rm th}\sigma_{\rm ann}v_\chi}\,,
\eea
where the thermalized DM number density is
\bea
n_\chi^{\rm th} = \frac{M_{\rm sg}/m_\chi}{(4\pi/3) r_{\rm th}^3} = \frac{\rho_c}{m_\chi}\,.
\eea
Requiring $t_{\rm ann} \gtrsim 1\,{\rm Gyr}$ yields a lower bound on the DM mass to ensure late-onset annihilation
\begin{align}
m_\chi \gtrsim 1.3 \times 10^9\,{\rm GeV} \,  \rho_6^{\frac{1}{3}} \left( \frac{\alpha_\chi}{0.1} \right)^{\frac{2}{3}} \,.
\label{eq:DMannCondition1}
\end{align}
This implies that, in contrast to the WIMP DM case, the annihilation of DMs with  $m_\chi \gtrsim 10^9\,{\rm GeV}$ typically remains inefficient even as its density grows large enough to induce gravitational collapse.

After collapse, the evolution of the DM core is governed by the total energy change of DM particles, combining contributions from scattering and annihilation, as follows
\bea
\frac{dE_\chi}{dt} =  \left. \frac{dE}{dt}\right|_{\rm scat} + \left. \frac{dE}{dt}\right|_{\rm ann} \, .
\label{eq:DMheatEQ1}
\eea
For a homogeneous DM distribution within the core, the virial theorem relates the average kinetic and potential energies of a DM particle
\bea
2\left<K_\chi\right> = - \left<V_\chi\right>\
\eea
with
\bea
\left<K_\chi\right>  =  \frac{1}{2}m_\chi \bar{v}_\chi^2 \,, \quad
\left<V_\chi\right>  = -\frac{3}{5}\frac{GN_\chi m_\chi^2}{r_\chi} \,,
\eea
where $\bar{v}_\chi$ is the average DM velocity, and $r_\chi$ is the characteristic size of the DM core.
The total DM energy is then
\bea
E_\chi = N_\chi \left(\left<K_\chi\right>+\left<V_\chi\right>\right) \simeq \frac{1}{2}N_\chi \left<V_\chi\right>\,,
\label{eq:DMvirial}
\eea
where $N_\chi$ is the total number of DM particles.

The energy loss rate due to elastic scatterings between DM and nuclei is estimated as
\begin{equation}\label{eq-dEdtsca}
\begin{split}
\left. \frac{dE}{dt}\right|_{\rm scat} & \simeq  - N_\chi \left<K_\chi\right>  \frac{\beta}{2} \, t_{\rm scat}^{-1} \\
& \simeq \frac{1}{2} E_\chi \beta \left(n_{\rm He} \sigma_{\rm \chi {\rm He}} v_{\rm rel}\right) \, .
\end{split}
\end{equation}
The second term on the right-hand side of \eqref{eq:DMheatEQ1} accounts for the total DM energy change through the DM pair annihilation.
Considering that each annihilation event removes two (identical) particles and releases their combined energy, we find
\begin{equation}\label{eq-dEdtann}
\begin{split}
\left. \frac{dE}{dt}\right|_{\rm ann} & \simeq  \frac{d N_\chi}{dt} \left( \left<K_\chi\right>+2\left<V_\chi\right>\right) \\
& \simeq - 3 E_\chi \left(n_\chi \sigma_{\rm ann} v_\chi\right) \, ,
\end{split}
\end{equation}
where in the second line we use the relation of
\bea
\frac{1}{N_\chi}\frac{dN_\chi}{dt} = - n_\chi \sigma_{\rm ann}v_\chi \, .
\eea

In the absence of annihilation, the total number of DM particles is conserved, and the total DM heating rate is governed solely by elastic scatterings.
This heating rate is maximized when the DM velocity reaches the decoherence threshold, $v_{\rm cut} 
\approx 0.02\,c$, as described in Sec.~\ref{sec:self-grav-collapse}.
The maximum elastic heating rate is given by
\begin{equation}
\begin{split}
    \dot{Q}_{\rm scat}  & = -\left[\left. \frac{dE}{dt}\right|_{\rm scat}\right]_{\bar{v}_\chi=v_{\rm cut}} \\
    & \simeq \frac{1}{4}  M_{\rm sg}  v_{\rm cut}^2   \beta \left(n_{\rm He} \sigma_{\rm \chi {\rm He}} v_{\rm cut}\right) \, ,
\end{split}
\label{eq:Qsca}
\end{equation}
where the prefactor indicates the mean DM energy loss per DM-He collision, and the term in the parentheses corresponds to the DM-He elastic scattering rate.
The corresponding DM number density at this point reads
\bea
\left[n_\chi\right]_{\rm cut} = n_\chi^{\rm th}\left(\frac{m_\chi v_{\rm cut}^2}{3T_c}\right)^3\,.
\label{eq:DMmaxDensity_elastic}
\eea
The total contraction timescale is approximately given by $2/\left(n_{\rm He} \sigma_{\rm \chi {\rm He}} v_{\rm cut}\right)$, which is comparable to the thermalization timescale $t_{\rm th}$ in \eqref{eq:thermaltimescale}.

In a more realistic scenario,
DM self-annihilation becomes significant as the density increases due to gravitational contraction.
Taking the derivative of \eqref{eq:DMvirial}, we have
\begin{align}
    \frac{1}{E_\chi} \frac{dE_\chi}{dt} &= \frac{1}{2N_\chi} \frac{dN_\chi}{dt} \langle V_\chi\rangle \nonumber\\
    &+\frac{1}{2} N_\chi \left( \frac{\partial \langle V_\chi\rangle}{\partial N_\chi} \frac{dN_\chi}{dt} + \frac{\partial \langle V_\chi\rangle}{\partial r_\chi} \frac{dr_\chi}{dt}\right) ~, 
\end{align}
which allows us to determine the evolution of $ r_\chi $. 
Equating $ dE_\chi/dt $ to the sum of \eqref{eq-dEdtsca} and \eqref{eq-dEdtann}, we arrive at
\bea
-\frac{1}{r_\chi}\frac{dr_\chi}{dt} = \frac{\beta}{2}\left(n_c \sigma_{\rm \chi {\rm He}} v_{\rm rel}\right) - n_\chi \sigma_{\rm ann} v_\chi  \, .
\eea
This yields the evolution equation for the DM number density
\begin{equation}
\begin{split}
    \frac{1}{n_\chi}\frac{dn_\chi}{dt} & = \frac{1}{N_\chi}\frac{dN_\chi}{dt}-\frac{3}{r_\chi}\frac{dr_\chi}{dt} \\
    & \simeq \frac{3}{2}\beta n_c \sigma_{\rm \chi {\rm He}} v_{\rm rel} - 4  n_\chi \sigma_{\rm ann} v_\chi \,.
\end{split}
\end{equation}
As a result, the DM number density increases until it reaches a saturation point (i.e., $dn_\chi/dt=0 $), where the contraction due to elastic scattering is balanced by annihilation-induced depletion
\bea
\left[n_\chi\right]_{\rm sat} = \frac{3 \beta n_c \,\sigma_{\rm \chi {\rm He}} v_{\rm rel}}{8\,\sigma_{\rm ann} v_\chi} \,.
\eea

This saturation can occur if $[n_\chi]_{\rm cut} > [n_\chi]_{\rm sat}$, which is generally satisfied more easily than the constraint in \eqref{eq:DMannCondition1} to prohibit pre-collapse annihilation. 
Specifically, this condition gives
\begin{align}
    m_\chi \gtrsim 3\times 10^{-14} \,{\rm GeV} \left( \frac{\sigma_{\chi{\rm He}}}{10^{-37}{\rm cm}^2} \right)  \left( \frac{v_{\rm rel}}{v_{\rm th}} \right) \left( \frac{\alpha_\chi}{0.1} \right)^2 ~,
\end{align}
where we have assumed the thermal velocity of helium nuclei dominates the relative velocity as a conservative estimation. 
Clearly, this condition is much easier to satisfy than~\eqref{eq:DMannCondition1}, so the saturation point is always reached in the parameter space of our interest if we turn on the annihilation.
Since annihilation becomes significant near the saturation point, where the total DM number $N_\chi$ begins to diminish exponentially, the maximal DM heating rate from annihilation, releasing the DM rest mass energy, is approximately given by
\begin{equation}
\begin{split}
    \dot{Q}_{\rm ann}  & \simeq m_\chi N_{\rm sg} \left[n_\chi\right]_{\rm sat} \sigma_{\rm ann}v_\chi \\
    & = \frac{3}{8} M_{\rm sg}  \beta n_c \,\sigma_{\rm \chi {\rm He}} v_{\rm rel} \, .
\end{split}
\label{eq:Qann}
\end{equation}
This rate significantly exceeds the maximum elastic heating rate in \eqref{eq:Qsca} (in the absence of DM self-annihilation). 
The ratio between them can be found as
\begin{align}
    \frac{\dot{Q}_{\rm ann}}{\dot{Q}_{\rm scat}} = \frac{3}{2} \frac{v_{\rm rel}}{v_{\rm cut}} \frac{1}{v_{\rm cut}^2} 
    = \mathcal{O}(10^3) ~,
\end{align}
where again we have conservatively taken $ v_{\rm rel} = v_{\rm th} $. 
This leads to a significant enhancement of the heating efficiency and, therefore, a much stronger constraint from the annihilation of DM particles than that with only elastic scatterings.

The localized overheating caused by energy deposition from collapsed DM particles can be redistributed via thermal diffusion into the surrounding cooler regions.
As DM heating continues, this process gradually raises the temperature of nearby material until the heating and diffusion rates are balanced.
For a diffusion region of size $r$,
the diffusion rate reads 
\begin{equation}
\begin{split}
    \dot{Q}_{\rm diff} & = - \frac{16\pi r^2 c}{3\kappa \rho_c} a T_c^3 \frac{d T}{d r} \\
    &\simeq -\frac{16\pi^3 r}{45} \lambda_{\rm eff} T_c^3 \left(T_b - T_c\right) \,,
\end{split}
\end{equation}
where $a= 4\sigma/c$ is the radiation constant with the Stefan-Boltzmann constant $\sigma$, and $\kappa$ denotes the opacity of the conducting medium with the relation of $ (\kappa \rho_c)^{-1} \simeq \lambda_{\rm eff} $.

If the heating rate $\dot{Q}_\chi$ exceeds the diffusion rate $|\dot{Q}_{\rm diff}|$, helium material within radius $r$ can be heated above $T_b$.
As discussed in Sec.~\ref{sec:TriggerMass}, if the heated region exceeds the volume of the corresponding trigger mass required for helium ignition at temperature $T_b$, this imbalance results in a runaway burning process, ultimately igniting the entire stellar core.

To quantify this condition, we adopt a benchmark burning temperature of $T_b = 6\times 10^8\,{\rm K}$, for which the critical mass condition is nearly saturated as shown in Fig.~\ref{fig:TriggerM}; see Fig.~\ref{fig:TriggerMintime} for the trigger mass at a respective time. 
The corresponding trigger radius $ r_{\rm tr} \simeq 62 {\rm cm} $ is significantly larger than the typical size of the collapsed DM core around $r_{\rm cut} \simeq 1.7\times 10^{-8}\,{\rm cm}$.
Therefore, our diffusion-based ignition criterion is self-consistent: the DM heating can be considered point-like relative to the ignition region of interest.
The diffusion rate at this threshold scale determines the critical DM heating rate for successful helium ignition.
Imposing that DM heating does not exceed this critical diffusion rate, i.e., $\dot{Q}_\chi < |\dot{Q}_{\rm diff}|_{r=r_{\rm tr}}$, places an upper bound on the DM–nucleus interaction cross section.

\subsection{Results}

\begin{figure}[t!]
    \centering
    \includegraphics[width=0.45\textwidth]{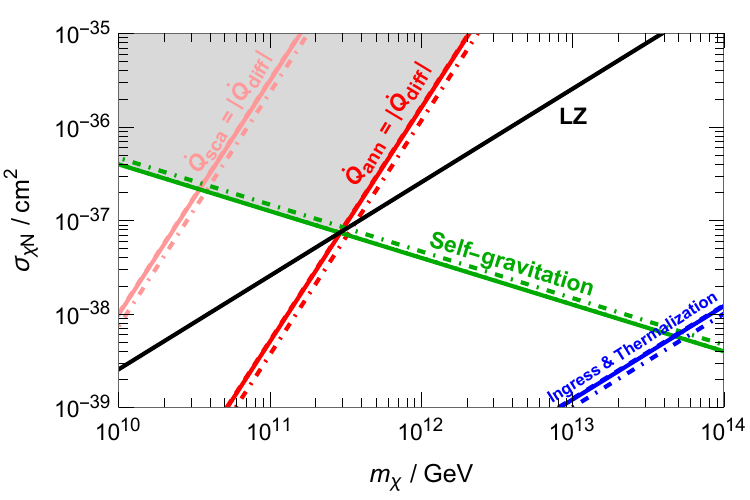}
    \caption{ Constraints on the spin-independent DM-nucleon cross section.
    The red (annihilation heating) and pale red (elastic-scattering heating only) curves show the ignition boundaries, defined by our runaway condition in the heated region. 
    The blue curve marks the combined requirement that captured DM can ingress to the core and thermalize within the relevant stellar timescale.
    The green curve indicates the threshold above which the captured DM becomes self-gravitating.
    The gray-shaded region is excluded: it corresponds to parameter points that simultaneously satisfy the capture and transport requirements (capture + ingress + thermalization), reach self-gravitation, and exceed the corresponding ignition boundary (pale red), implying a DM-induced premature helium flash under our adopted stellar profile and localized-heating treatment.
   Solid curves correspond to the benchmark model with $(M, Z) = (0.83\,M_\odot, 0.001)$. 
    For comparison, the region above the black line is excluded by terrestrial direct detection experiments \cite{LZ:2022lsv}. 
    To confirm the robustness of our results, we also show results obtained using alternative stellar-evolution profiles: $(0.83\,M_\odot, 0.0001)$ (dot-dashed) and $(0.9\,M_\odot, 0.001)$ (dashed).
    }
    \label{fig:result}
\end{figure}

In Fig.~\ref{fig:result}, the red lines indicate the DM parameter space, where the DM heating rate can exceed the diffusion rate that allows for runaway helium ignition in the stellar core.
The pale red line ($\dot{Q}_{\rm sca} = |\dot{Q}_{\rm diff}|$) corresponds to the case without DM annihilation, where heating arises solely from DM-nucleus elastic scattering (i.e, kinetic heating), as given by \eqref{eq:Qsca}.
The solid red line ($\dot{Q}_{\rm ann} = |\dot{Q}_{\rm diff}|$) includes the additional contribution from DM self-annihilation, described by \eqref{eq:Qann}, which enhances the heating rate by roughly three orders of magnitude as discussed above.

However, the red contours indicating successful DM-induced heating are only valid if all prerequisite conditions for DM ingress, thermalization, and collapse (outlined in Sec.~\ref{sec:DM_capture}) are satisfied.
The conditions for ingress and thermalization can be formulated in terms of characteristic timescales, each of which must be shorter than the RG lifetime, approximately $t_{\rm RG} \approx 1\,\mathrm{Gyr}$:
\begin{itemize}
    \item $t_{\rm ingress} < t_{\rm RG}$: After being captured, DM particles gradually lose energy through repeated scatterings with nuclei. Ingress is complete when their orbits become fully confined within the RG core. This must occur early enough to enable subsequent evolution. 
    \item $t_{\rm th} < t_{\rm RG}$: Once confined, DM particles continue to scatter until their kinetic energy becomes comparable to the core temperature. The thermalization process must be completed within the RG lifetime.
\end{itemize}
The other condition for DM collapse reads
\begin{itemize}
    \item $N_\chi (t_{\rm end}) > M_{\rm sg}/m_\chi$: After thermalization, the DM distribution remains extended until its density exceeds that of the ambient stellar medium. At this point, DM self-gravity dominates, and the DM core collapses. For this to contribute to helium ignition, the collapse must occur within the stellar lifetime.
\end{itemize}
Here, as discussed in Sec.~\ref{sec:numerics}, $t_{\rm end}$ denotes the truncated evolution time, which is chosen to ensure that a premature helium flash results in a luminosity sufficiently lower than the TRGB value, accounting for theoretical and observational uncertainties.
We note that for cross section larger than this threshold, DM collapse can occur earlier than $t_{\rm end}$, but the corresponding heating-cooling competition condition is typically satisfied without difficulty.
Furthermore, due to the significantly enhanced capture efficiency during the RG phase, additional DMs captured and accumulated later in the evolution can also readily fulfil this collapse condition.

In Fig.~\ref{fig:result}, the blue and green lines represent the boundaries imposed by the combined ingress and thermalization conditions, and the self-gravitation criterion, respectively.
The region above all these lines defines the DM parameter space in which capture, collapse, and subsequent heating can occur within the relevant RG evolutionary timescale, enabling the subsequent heating required for helium ignition.

As shown in Fig.~\ref{fig:result}, the ingress and thermalization conditions (blue curve) are easily satisfied across a broad range of DM parameters and thus impose only mild constraints. 
The self-gravitation condition yields a more stringent bound, with the green contour representing the critical line above which DM particles can undergo collapse within the RG lifetime.
In the absence of annihilation, elastic scattering alone can provide sufficient heating to ignite helium, but the corresponding parameter region (pale red curve) lies above the exclusion limit from current direct detection experiments.
However, when DM self-annihilation is involved, the heating rate is significantly enhanced, shifting the viable region downward in the parameter space (red curve).
This accounts for a significant sensitivity of RGs in parameter space for HDM, in particular with masses around $10^{11}\,{\rm GeV}$ and spin-independent nucleon cross sections on the order of $10^{-37}\,{\rm cm}^2$ (gray-shaded).
This result highlights a novel astrophysical sensitivity to HDM candidates comparable to the reach of existing terrestrial searches.

As a remark, to evaluate the sensitivity of our results to the underlying stellar model, we repeat the full analysis using additional RG profiles with different initial masses and metallicities, chosen to remain consistent with the ages and metallicities of the halo globular clusters relevant for TRGB-based constraints.
In particular, besides our benchmark model $(M\,,Z) = (0.83\,M_\odot\,,0.001)$ illustrated by solid curves, which serves as a representative case, we take into account $(0.83\,M_\odot\,,0.0001)$ to probe lower metallicity (dot-dashed) and $(0.9\,M_\odot\,,0.001)$ to probe a modest increase in mass corresponding to an younger $(\sim 11\,{\rm Gyr})$ population (dashed).
We realize that the resulting exclusion contours are nearly unchanged and largely overlap those of the benchmark case, which indicates that our results are robust against realistic variations in mass and metallicity within the parameter range relevant to halo globular clusters.
In practice, the residual uncertainty associated with these stellar-profile variations is small compared with other theoretical uncertainties, most notably those related to the DM capture and subsequent collapse/heating phenomenology.

\section{Conclusion and Discussion}

In this work, we have investigated the possibility that captured DMs inside RG cores can trigger premature helium ignition via localized heating.
Such an event would alter the stellar evolutionary path of low-mass stars and their observational consequences, particularly lowering the luminosity at the TRGB.
This provides a novel astrophysical probe of DM interactions with baryonic matter.

We began by analyzing DM capture via elastic scattering with nuclei in the RG core and identified the parametric regimes where the capture rate scales linearly with the scattering cross section and where it saturates at the geometric limit.
We derived analytic expressions for the capture rate that agree with the full numerical results across a broad range of DM masses and cross sections (Sec.~\ref{sec:capture}).
Once captured, the DM particles accumulate in the stellar core.
For most of the relevant parameter space, DM ingress (Sec.~\ref{sec:ingress}) and thermalization (Sec.~\ref{sec:thermalization}) occur on relatively short timescales, and the DM population becomes concentrated in a compact region.
When the total DM mass exceeds the self-gravitating threshold $M_{\rm sg}$, the DM core becomes unstable and collapses under its own gravity (Sec.~\ref{sec:self-grav-collapse}).
This collapse proceeds through a sequence of virialized configurations regulated by scatterings with the stellar medium.

This dense, compact DM core can serve as a localized heat source, either via annihilation or continued elastic scattering, which potentially raises the temperature of the surrounding material.
We estimated the critical trigger mass and radius for helium ignition by balancing the nuclear energy release against conductive diffusion losses (Sec.\ref{sec:TriggerMass}).
Using this framework, we developed a self-consistent diffusion-based ignition criterion to compare the DM heating rate to the diffusive losses over the relevant spatial scale (Sec.\ref{sec:DMheatingVSdiffusion}).

The resulting constraints on DM parameter space are summarized in Fig.~\ref{fig:result}.
Ingress and thermalization conditions (magenta curve) impose only mild restrictions, while the self-gravitating collapse sets a stronger bound (green curve).
In the absence of annihilation, elastic scattering can still provide sufficient heating (red curve), though this region is already excluded by current direct detection bounds.
However, when annihilation is included, the heating rate is significantly enhanced by a factor of $1/v_{\rm cut}^2 \sim 10^3$, that leads to a sensitive probe of DM as comparable to direct detection searches, particularly for masses around $10^{11}\,{\rm GeV}$ and spin-independent cross sections around $10^{-37}\,{\rm cm}^2$.
We note that the current best direct detection constraint~\cite{LZ:2022lsv} in Fig.~\ref{fig:result} is extrapolated from the constraint for the sub-$\rm TeV$ DM and may be less robust in the high mass, large cross section regime due to potential Earth-shielding effects.
Hence, RG observations offer a complementary probe of DM in this otherwise inaccessible regime.

It is noteworthy to compare our results with earlier studies~\cite{Lopes:2021jcy,Dessert:2021wjx,Hong:2024ozz} that also explore DM-triggered helium flashes in RGs.
Refs.~\cite{Lopes:2021jcy} and \cite{Hong:2024ozz} focus on weak scale DM, $m_\chi\sim 1\,{\rm TeV}$, for which annihilation can become important even before gravitational collapse.
In that regime, energy deposition from DM self-annihilation, which is equilibrated with the capture rate, occurs over spatial scales much larger than the thermal radius scale, which implies a heated region that is far more extended than the compact, near point-like heating relevant to our setup.
Consequently, analyses in this mass range generally require a broader hydrodynamical treatment of the stellar response.
Ref.~\cite{Dessert:2021wjx} examines macroscopic DM with masses of $10^{17}$-$10^{20}\,{\rm g}$, which also leads to heating over large spatial scales comparable to a non-gravitational interaction size.
While all these studies, including ours, apply similar diffusion-based runaway criteria, our analysis is distinct in exploring helium ignition under much smaller ignition radii, relevant for DM in the mass regime, where pre-collapse annihilation is negligible.
This allows for a significant localized heating scenario and a correspondingly more stringent ignition condition.

In conclusion, RGB stars provide a sensitive and complementary probe of heavy DM with sizable interactions.
Our findings highlight a novel astrophysical window into DM that could be further explored with improved stellar modeling and future observational advances.

\section{Acknowledgment}

We thank Andrea Caputo, Hyungjin Kim, Jorge Martin-Camalich, and Meng-Ru Wu for useful discussions.
SG, MH, CSS, and SY are supported by IBS under the project code, IBS-R018-D1.
CSS is supported by Global-Learning \& Academic research institution for Master’s·PhD students, and Postdocs (G-LAMP) Program of the National Research Foundation of Korea (NRF) grant funded by the Ministry of Education (No. RS-2025-25442707), and by the NRF grant funded by the Korean government (MSIT) (No. RS-2026-25498521). O.S. acknowledges the support of the large-program BRAVOSUN funded by the Italian National Institute of Astrophysics.

\appendix






\section{Approximated derivation of $ g_N $}
\label{sec:gN}

In this appendix, we show that the capture probability $ g_N $ can be evaluated analytically under a linear approximation, which is valid for $ \beta N \ll 1 $.
Here, $\beta$ denotes the parameter for the fractional energy loss per DM-nucleus scattering, defined in \eqref{eq:energy_loss}, and $N$ is the number of scatterings.

From the definition in \eqref{eq:capture_prob}, we can expand the product inside the step function to first order in $ \beta $ as 
\begin{equation}
    \prod_{i=1}^N (1-\beta x_i) \simeq  1- \beta \sum_{i=1}^N x_i \,.
\end{equation}
The capture probability then approximates to
\begin{equation}
     g_N \simeq \left[\prod_{i = 1}^N\int_0^1 d x_i \, f_i ( x_i) \right] \Theta \left(-\frac{1}{\beta}\frac{u^2}{\omega^2} + y_N\right)\,,
\end{equation}
where we define
\begin{equation}
    y_N \equiv \sum_{i=1}^N x_i \,.
\end{equation}

Assuming identical and independent distributions for all $ x_i $, such that $f_i(x_i)=f(x_i)$, we can apply the central limit theorem in the large $N\gg 1$ limit.
In this case, the distribution of $ y_N$ approaches a Gaussian 
\begin{align}
    g_N(\omega, \beta) \simeq \int_{0}^{N} dy_N \,f_{y_N} (y_N) \Theta \left(-\frac{1}{\beta}\frac{u^2}{\omega^2} + \, y_N \right) \,,
\end{align}
where $f_{y_N}$ is the probability distribution of $y_N$, approximated by a Gaussian with mean $ N \times {\rm E}[x_i] $ and variance $ N \times {\rm Var}[x_i] $ with ${\rm E}[x_i]$ and ${\rm Var}[x_i]$ denoting the mean and variance of $x_i$, respectively.

For a uniform distribution of $x_i \in [0,1] $, i.e., $f(x_i) = 1$, we have ${\rm E}[x_i]=1/2 $ and $ {\rm Var}[x_i]=1/12 $.
This results in 
\begin{align}
    f_{Y_N} (y_N) \simeq \frac{1}{\sqrt{2\pi}\sqrt{N/12}}  \exp \left[ - \frac{\left( y_N -N/2 \right)^2  }{N/6} \right] \,.
\end{align}
The capture probability can be written analytically in terms of the error function (${\rm Erf}$) as
\begin{align}
    g_N \simeq \frac{1}{2} \left( {\rm Erf} \left[\sqrt{\frac{3N}{2}}\right] - {\rm Erf} \left[\frac{ \frac{u^2}{\beta \, \omega^2} -\frac{N}{2}}{\sqrt{N/6}}\right] \right) \,. 
\end{align}

\section{Mean free paths for radiative and conductive diffusion in red giant cores}
\label{app:mfp}

In this appendix, we derive the mean free paths of photons and electrons, which are the primary carriers of energy through radiative and conductive diffusion, respectively, in helium-rich cores of RG stars.

Photons in RG cores interact through scattering with free electrons and via inverse bremsstrahlung absorption (free-free absorption).
The former is described by Thomson scattering in the non-relativistic limit, leading to the following opacity~\cite{Shapiro:1983du}
\begin{align}
    \kappa_{{\rm f}\gamma} = \frac{n_e}{\rho_c} \sigma_{\rm Th} f_\eta \simeq 0.4 \left(\frac{Z}{A}\right) f_\eta \,{\rm cm}^2\, {\rm g}^{-1} \,,
\end{align}
where $A =4$ and $Z =2$ are the atomic and charge numbers of helium nuclei, respectively.
The function $f_\eta$ accounts for suppression due to electron degeneracy~\cite{1968stph.book.....C,1976ApJ...210..440B,Weaver:1978zz,1983ApJ...267..315P,Poutanen:2016kqx}, for which we adopt the analytic fit from Ref.~\cite{Poutanen:2016kqx}.

For inverse bremsstrahlung absorption in the low-temperature regime (where positrons are negligible), the Rosseland mean opacity is given by~\cite{1985ApJ...294...17I}
\begin{align}
	\kappa_\text{ff} = \left[ \frac{15}{4\pi^4} \rho_c \int_0^\infty \frac{u^4 e^{-u}du}{(1-e^{-u})^3 n_e \langle \sigma^- \rangle} \right]^{-1} \,,
\end{align}
where $ \langle \sigma^- \rangle $ the thermally averaged cross section for inverse bremsstrahlung.
The following relation connects this cross section to temperature and density:
\begin{align}
	\frac{n_e \langle \sigma^- \rangle}{\rho_c} = 7.4\times 10^{2} \,{\rm cm}^2\,{\rm g}^{-1} \, \rho_6 T_8^{-\frac{7}{2}} \frac{g_\text{He}}{u^3} \,,
\end{align}
where $g_{\rm He} \sim \mathcal{O}(1)$ is the Gaunt factor~\cite{1985ApJ...294...17I,1961ApJS....6..167K}.
This leads to an opacity that obeys the Kramers law in terms of the density and temperature.
The total photon mean free path is then given by
\bea
    \lambda_r = \frac{1}{\rho\left(\kappa_{{\rm f}\gamma}+\kappa_{\rm ff}\right)} \,.
    \label{eq:lambdaphoton}
\eea

The conductive opacity is expressed as
\bea
    \kappa_e = \frac{16\sigma T^3}{3\rho \xi}\,,
    \label{eq:eOP}
\eea
where $\xi$ is the thermal conductivity.
In the degenerate regime, $\xi$ takes the form
\bea
    \xi = \frac{\pi^2 n_e T}{3m_e \nu_e}\,
\eea
with $\nu_e$ the total electron scattering frequency.
This includes contributions from both electron-ion (ei) and electron-electron (ee) scattering.

The frequency for electron-ion (Mott) scattering is~\cite{1980SvA....24..303Y}
\begin{align}
	\nu_{ei}=  \frac{4\pi Z^2 \alpha^2}{p_{\rm F}^2 v_{\rm F}}  \, n_i \, \Lambda_{ei} \,,
\end{align}
where $p_{\rm F}$ is the electron Fermi momentum, $v_{\rm F} = p_{\rm F}/\sqrt{p_{\rm F}^2+m_e^2}$ is the Fermi velocity, and $n_i $ is the ion number density.
The Coulomb logarithm $\Lambda_{ei}$ is given by
\begin{equation}
	\Lambda_{ei}  =  \ln \left[ \left( \frac{2\pi Z}{3} \right)^{\frac{1}{3}} \left( 1.5+ \frac{3}{\Gamma} \right)^{\frac{1}{2}} \right] -\frac{v_{\rm F}^2}{2} \,
\end{equation}
with the Coulomb plasma parameter defined as
\begin{align}
	\Gamma = \frac{Z^2\alpha}{T} \left( \frac{4\pi n_i}{3} \right)^{1/3} \,.
\end{align}

The electron-electron scattering frequency~\cite{Lampe:1968zz,1980SvA....24..126U} incorporates plasma screening effects and follows the fitted expression~\cite{Shternin:2006uq,Cassisi:2007ty}
\begin{align}
	\nu_{ee} = \frac{6\alpha^{\frac{3}{2}}}{\pi^{\frac{5}{2}}}\,p_{\rm F}\,\sqrt{v_{\rm F}}\,y\, I\left(v_{\rm F},y\right)
\end{align}
where $ y\equiv \sqrt{3} \omega_{\rm pl}/T $, with $ \omega_{\rm pl} =\sqrt{4\pi\alpha n_e/(p_{\rm F}^2+m_e^2)^{1/2}}$ the plasma frequency, and~\cite{Shternin:2006uq,Cassisi:2007ty}
\begin{equation}
    \begin{split}
        I(v_{\rm F},y) & = \frac{1}{v_{\rm F}}\left(\frac{10}{63}-\frac{8/315}{1+0.0435 y}\right) \\
        &\times \ln\left[1+\frac{128.56}{37.1 y+10.83 y^2+y^3}\right]\\
        + & v_{\rm F}^3\left(\frac{2.404}{\mathtt{B}}+\frac{\mathtt{C}-2.404/\mathtt{B}}{1+0.1v_{\rm F}y}\right)\\
        &\times \ln\left[1+\frac{\mathtt{B}}{\mathtt{A}v_{\rm F}y+v_{\rm F}^2y^2}\right]\\
        + & \frac{v_{\rm F}}{1+\mathtt{D}}\left(\mathtt{C}+\frac{18.52v_{\rm F}^2 \mathtt{D}}{\mathtt{B}}\right)\\
        &\times \ln\left[1+\frac{\mathtt{B}}{\mathtt{A}y+10.83v_{\rm F}^2y^2+v_{\rm F}^{8/3}y^{8/3}}\right]
    \end{split}
\end{equation}
with $\mathtt{A} = 12.2+25.2v_{\rm F}^3$, $\mathtt{B}=\mathtt{A}\exp[(0.123636+0.016234v_{\rm F}^2)/\mathtt{C}]$, $\mathtt{C}=8/105+0.05714v_{\rm F}^4$, and $\mathtt{D}=0.1558y^{1-0.75v_{\rm F}}$.
Summing both contributions,
\begin{align}
\nu_e = \nu_{ei} + \nu_{ee} \,,
\end{align}
we can evaluate $\kappa_e$ using \eqref{eq:eOP}, and define the electron mean free path as
\begin{align}
\lambda_e = \frac{1}{\rho \kappa_e} \,.
\end{align}

Finally, the effective mean free path governing energy diffusion in the RG core is given by the sum of the radiative and conductive contributions
\begin{align}
\lambda_{\rm eff} = \lambda_r + \lambda_e \,.
\end{align}

 \bibliographystyle{JHEP}
    \bibliography{refs}
\end{document}